          [1999/12/02 1.01 (PWD)]
\documentclass[a4paper,twocolumn]{esapub} 

\def\ltsima{$\; \buildrel < \over \sim \;$}
\def\lsim{\lower.5ex\hbox{\ltsima}}
\def\gtsima{$\; \buildrel > \over \sim \;$}
\def\gsim{\lower.5ex\hbox{\gtsima}}

\usepackage{epsfig}

\title{The INTEGRAL Burst Alert System}

\author[1]{S. Mereghetti}
\author[1]{D.I. Cremonesi}
\author[2]{J. Borkowski}

\affil[1]{Istituto di Fisica Cosmica ''G.Occhialini'' -- CNR,
Milano, Italy} \affil[2]{INTEGRAL Science Data Center, Versoix,
Switzerland}

\begin{document}


\maketitle

\begin{abstract}
We describe the  INTEGRAL Burst Alert System (IBAS): the automatic software developed
at the INTEGRAL Science Data Center to allow the rapid distribution of the coordinates
of the Gamma-Ray Bursts detected by INTEGRAL.

IBAS is implemented as a ground based system, working on the
near-real time telemetry stream. It is expected that the system
will detect more than one GRB per month in the field of view of
the main instruments. Positions with an accuracy of a few
arcminutes will be distributed to the community for follow-up
observations within a few tens of seconds of the event. The
system will also upload commands to optimize the possible
detection of   bursts in the visible band with the INTEGRAL
Optical Monitor Camera.
\end{abstract}

\section{Introduction}

The origin of gamma-ray bursts (GRB's) remained one of the great mysteries
of high energy astrophysics for more than 25 years after the publication
of their discovery (Klebesadel et al. 1973).
The puzzle became even more complicated after the first results
obtained with the BATSE instrument. This detector, especially designed
to significantly increase the GRB sample, led to the unexpected result of a
non-euclidean LogN-LogS coupled with a uniform angular distribution in
the sky (see, e.g., Fishman 1995, and references therein).
It soon became clear that the only way to make a significant advance in
the field was the possibility of identifying the counterparts of GRB at other
wavelengths.

The instruments on board INTEGRAL, though not specifically
optimized to observe GRB's, offer the possibility of rapidly
obtaining accurate positions of the GRB's observed by chance in
their large field of view. It was therefore proposed to implement
a ''burst alert system'' in order to allow rapid
multi-wavelength  follow-ups (Pedersen et al. 1997). Such a
proposal was boosted by the exciting results obtained with the
\textit{BeppoSAX} satellite (Costa et al. 1997, van Paradijs et
al. 1997), that clearly demonstrated the capabilities of a coded
mask instrument, similar to the INTEGRAL ones, in quickly
localizing GRB's.

\section{Overview of the INTEGRAL Burst Alert System (IBAS) }

Since no on-board GRB detection system is  foreseen on  INTEGRAL,
the search for GRB's   will  be performed on   ground by means of
a near real time analysis running at the   INTEGRAL
Science Data Center (ISDC).
The telemetry data, received at the MOC, will be transmitted on a 128 kbs dedicated
line to the ISDC. Here the relevant data packets will be extracted and
immediately fed into a dedicated software system called
IBAS (Integral Burst Alert System), independent from the main
data processing pipeline of the ISDC (Mereghetti et al. 1999).


It is expected that ISGRI, the  first layer of the IBIS instrument,
operating in the 15-300 keV energy range, will yield the best
chances to detect a large number of GRB's and to accurately determine
their positions.
The SPI instrument, with a similar sensitivity and large field of view, can also
detect GRB's, but its angular resolution is not as good as the IBIS one.
Relatively strong bursts in the central part of
the field of view will be detectable by   SPI and IBIS, therefore a simultaneous
trigger in both instruments could be used to increase the confidence in the reality of
the event. However, due to the different  sensitivity curves as a function of
energy and source direction,  it is   likely
that several GRB's  will   trigger   only  in a single detector.
For this reason the IBAS validation process will not require a multiple detection
to confirm the occurrence of a GRB.

The first step of the GRB search will be based on a simple monitoring of the incoming
count rates, without resorting to more complex image analysis.
In practice this will be done by  looking for  significant excesses with respect to
a running average, in a way similar to traditional on-board triggering algorithms.
In fact any transient source strong enough to appear as a significant new peak in the deconvolved
images will also produce a detectable excess in the overall count rate.
The search will be simultaneously performed in many different time scales and
energy ranges, to optimize the sensitivity to GRB's with different characteristics.

\begin{figure}[ht] 
\centerline{\epsfig{file=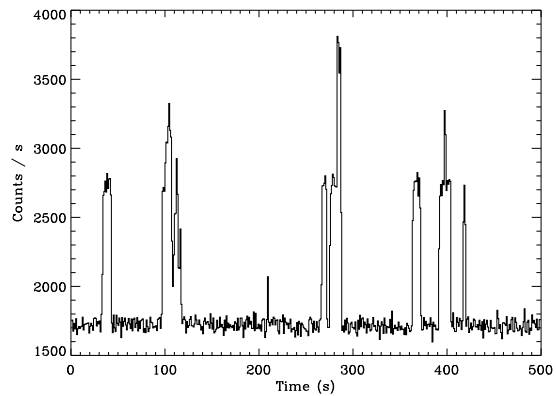,height=3.in,width=2.2in,
angle=90}} \vspace{10pt} \caption{Example of simulated ISGRI data containing
a GRB and several noisy pixels. The figure
shows the light curve of the overall count rate on the instrument, including the
background ($\sim$1200 cts s$^{-1}$),
a source with intensity of 1 Crab ($\sim$530 cts s$^{-1}$),
a GRB with equivalent peak flux in the 50-300 keV range of $\sim$0.7 cts cm$^{-2}$ s$^{-1}$,
and several noisy pixels producing short spikes of high count rate.
}
\label{lcurve}
\end{figure}

\begin{figure}[ht] 
\centerline{\epsfig{file=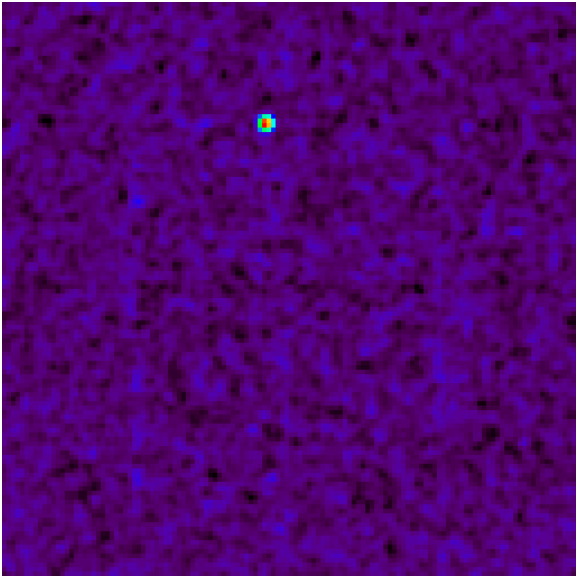,height=3.in,width=3.in,
angle=90}} \vspace{10pt} \caption{Image of the GRB event shown in
Figure 1. The reference image of a time interval preceding the trigger has been
subtracted, thus the GRB can be easily identified as  the only peak in the
deconvolved image. }
\label{grb}
\end{figure}

\begin{figure}[ht] %
\centerline{\epsfig{file=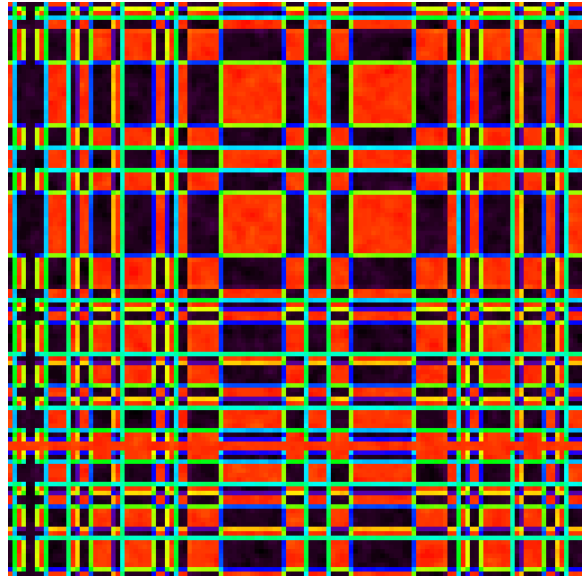,height=3.in,width=3.in,
angle=90}} \vspace{10pt} \caption{Deconvolved image corresponding to the
time interval of one of the noisy pixel spikes shown in Figure 1. The image pattern,
without a point source, clearly allows to
reject this trigger.
  }
\label{hotpix}
\end{figure}

When a candidate event is detected, a process of image analysis shall start
to verify the origin of the count rate variation and to ensure that the event
was not caused by an instrumental malfunctioning
(e.g. a bad detector pixel) or by a background variation
(see Figures  \ref{lcurve}, \ref{grb}, \ref{hotpix}).
Images shall be accumulated for different  time intervals,
deconvolved with very fast algorithms, and compared to the pre-burst
reference images in order to detect the appearance of the GRB as a new source.
This last step is of fundamental importance, since in general different sources
will be present in the field of view.
The image analysis will be based on the time intervals,
derived from the GRB light curve,  that optimize the
signal to noise ratio.

If the event is genuine,  the satellite attitude
information will be applied to derive a sky position
that is then automatically transmitted,
by e-mail and/or direct TCP/IP socket,  to all the subscribed users.
In addition, if the GRB is located in the sky region covered by the
OMC, an appropriate telecommand will be generated and sent to the satellite
to reconfigure its observing parameters (see Section 6).
Because full event validation and localization might require a
longer time, we foresee different levels of alert messages providing
increasingly accurate and reliable information.
These messages will be configured in such a way to allow an easy filtering
by the users in order to react only to the situations that best fit their needs.

\begin{figure}[ht] %
\centerline{\epsfig{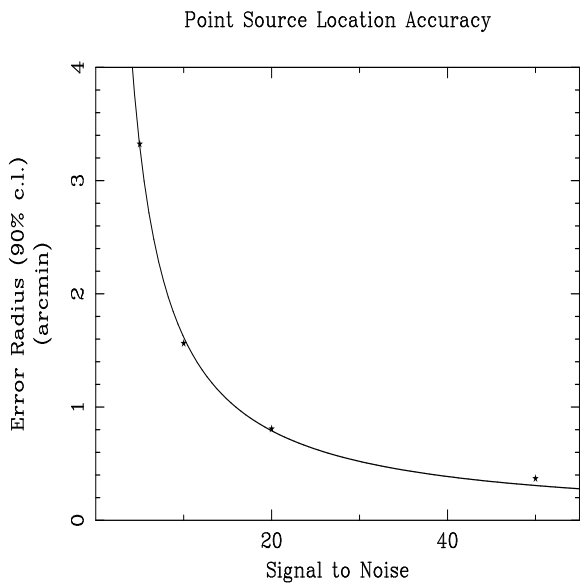}} \vspace{10pt} \caption{Source Location Accuracy as a
function of the Signal to Noise ratio for the ISGRI detector.
  }
\label{sla}
\end{figure}

\section{Number of expected GRB's}

A very simple estimate of the number of GRB's expected within the field
of view of the INTEGRAL instruments can be done by scaling the total
rate of events measured by BATSE ($\sim$666 GRB  year$^{-1}$,
Paciesas et al. 1999). For the field of view of IBIS
($\sim$30$^{\circ} \times$ 30$^{\circ}$),
this yields  666 $\times$ (0.23 sterad / 4$\pi$) $\sim$ 12 GRB year$^{-1}$.

A more accurate estimate must take into account the varying sensitivity within
the IBIS field of view and the different energy range of the BATSE detectors.
This can be done by convolving the IBIS sensitivity and solid angle as a function of
off-axis angle with the BATSE LogN-LogP relations, converted to the appropriate
energy range assuming an average GRB spectral shape. It turns out that such
computations do not significantly change the above rough result.
In fact we obtain   expected rates of $\sim$13, 10 and 8 GRB year$^{-1}$,
within the ISGRI field of view, adopting respectively
the LogN-LogP corresponding to BATSE trigger times $\Delta$T of 1 s, 256 ms and 64 ms.

The major uncertainty on the derived GRB rates is related to the extrapolation
of the BATSE LogN-LogP curves down to the ISGRI sensitivity ($\sim$0.1
ph cm$^{-2}$ s$^{-1}$, 50-300 keV peak flux for $\Delta$T=1 s).
Such an extrapolation depends on
the poorly known spectral shape of very weak GRB's.

It must be noted that these estimates do not take into account the likely existence of
GRB's with different characteristics than those observed by BATSE. For instance it is clear that
BATSE had a limited sensitivity for events shorter than 64 ms, as well as a
strong bias against the detection of long, slowly rising GRB's.
It it therefore very likely that the figures reported above will be an underestimate,
and that a few events per month will be available through IBAS for rapid multi-wavelength
follow-up  observations.

\section{Location Accuracy}

The source location accuracy (SLA) of coded mask imaging systems
depends on  the signal to noise ratio of the source.
For sources detected with a high statistical significance the
SLA can be a small fraction of the angular resolution. Theoretical
evaluations, confirmed by several independent simulations, have shown that
for ISGRI a SLA smaller than 30$''$ (90\% confidence level)
can be obtained for a signal to noise ratio of 30
(see Figure \ref{sla}).
In such cases, the final accuracy on the GRB location is also affected
by the uncertainties on the satellite attitude (see below).
Of course, most of the detected GRB's will have relatively small signal to noise
ratios, resulting in typical uncertainties, dominated by the photon statistics,
of the order of $\sim$2-3$'$.

The expected INTEGRAL attitude accuracy depends on three
cases: (1) attitude during slews,
(2) attitude at the start of a stable pointing period, and
(3) attitude from about ten minutes
 after the start to the end of a stable pointing period.

For case (1)  a slew path
simulator   generates predictions of the  pointing direction every
10 seconds.
In most circumstances these predictions should be
accurate to within 10$'$ (3 $\sigma$). The average slew rate
during the Galactic Plane scans and during the normal dithering observations
will be of the order of   $\sim$1$^{\circ}$/min.
Therefore,  it will be possible to detect
GRB's  even during slews, although their location accuracy will be worse.

For case (2) the attitude is based upon a prediction of the expected position
with respect to the previously commanded slew.
In  most cases (i.e. slews shorter than 2$^{\circ}$)
these predicted values will have
an accuracy of the order of the star-tracker/instrument alignment uncertainty
(30$''$ or less). For larger slews
the predicted attitude will have an error smaller than 5$\pm$1$'$
(3$\sigma$).
%

Approximately 5 minutes after the start of a pointing
(case 3)  a ``snapshot'' attitude
reconstruction will be completed   based
upon the first down-linked star-tracker map.
The result shall be   made available to the IBAS
within a maximum delay of 10 minutes since the start of the
stable pointing,
yielding, as above, an attitude with accuracy  $\lsim$30$''$.

\section{IBAS time performance}

Using simulated telemetry data, we have obtained the following
performance figures for the current prototype version of IBAS
running on a  SUN ULTRA 10   workstation
(440 MHz clock, 256 Mbytes RAM). Of course the final version will
run on a faster workstation exclusively dedicated to the IBAS system.

The first IBAS steps (telemetry receipt, extraction and sorting of the relevant
data packets, photon binning for the trigger search) require less than 0.1 s.
 
The speed of the triggering algorithm depends on the duration of the
smallest time interval considered and on the number of timescales.
For example, sampling on bins  of 1, 4, 16, 64 and 256 ms,
with a resolution of 1 ms it is still possible
to process the telemetry at a speed twice faster than
that of the real-time incoming data.

The most time consuming tasks are those related to the image analysis.
For this reason, the first part of imaging after a trigger
is based on a   detection algorithm able to discover
new sources by only considering (ghost) peaks in the fully coded
field of view. This can be done within $\sim$200 ms for ISGRI
and it allows to discriminate between false triggers and
true events. Only when a likely point source is detected, a
more thorough analysis will be done to locate the GRB. The deconvolution
of the whole (totally plus partially coded) field of view
and localization of the true source peak
currently requires $\sim$5 s (we expect to improve this performance
with a new optimized version of the code).

Thus it should be possible to send the first GRB alert within a few seconds
after the trigger time. The delay between the trigger time and the GRB onset
is of course dependent on the intensity and  time profile of the event,
but the IBAS simultaneous sampling in different timescales should ensure
a minimum delay in most cases.

To this time budget one has to add the time required for the telemetry
transmission from the satellite to the ISDC, that under normal  circumstances
should be smaller than $\sim$30 s.

Thus, in many cases, we foresee to be able to generate {\it first level} alerts
while the  GRB is still ongoing.

\section{Optical Monitor Recomanding}

The Optical Monitor Camera (OMC) on board INTEGRAL has a field of view
of 5$^{\circ} \times$5$^{\circ}$, but owing to the limited allocated telemetry
($\sim$1.7 kbit s$^{-1}$)
only the data from a number of preselected windows around sources of interest will be
downloaded during normal operations.
IBAS will check whether the derived GRB position falls within the OMC
field of view. In such a case, the appropriate telecommand with the definition
of a new window centered on the interesting region will be generated and sent to
the satellite.
This will allow to quickly observe the   GRB/afterglow emission in the optical band.
The OMC observation should consist of many frames with short integration times
to permit variability studies and to increase the sensitivity for very intense but
short outbursts. The expected limiting magnitude
is of the order of V$\sim$14-15 for an integration
time of 10 s, and  V$\sim$12-13 for 1 s.

\section*{Acknowledgments}
%
We are grateful to several people that have contributed to the IBAS project,
including G.Bazzaro, S.Brandt, D.Jennings, R.Hudec, H.Pedersen,
A.Pellizzoni, M.Pohl, T.Courvoisier and R.Walter.
The IBAS development    is supported by the Italian Space Agency.
JB was supported by the Polish Committee for Scientific Research
(KBN) under grant No. 2P03C00619p02.


\end{document}